\documentclass{aastex}

\begin{document}

\title{The EUV excess emission of the Virgo and 
A1795 clusters - re-observation with in-situ background measurements}

\author{Massimiliano~Bonamente$\,^{1}$, Richard~Lieu$\,^{2}$ and
Jonathan~P.~D.~Mittaz$\,^{3}$}

\affil{\(^{\scriptstyle 1} \){Osservatorio Astrofisico di Catania, Via S. Sofia 78, I-95125
Catania, Italy}\\
\(^{\scriptstyle 2} \){Department of Physics, University of Alabama,
Huntsville, AL 35899, U.S.A.}\\
\(^{\scriptstyle 3} \){Mullard Space Science Laboratory, UCL,
Holmbury St. Mary, Dorking, Surrey, RH5 6NT, U.K.}\\
}

\begin{abstract}
The Virgo and A1795 clusters of galaxies were re-observed
by EUVE with {\it in situ} background
measurements by pointing at small offsets.  Earlier, a similar
re-observational strategy applied to the cluster A2199 revealed
that the background radial profile was 
consistent with a flat distribution,  and therefore the
original method of extracting cluster EUV signals by the subtraction of
an aymptotically determined background was valid.
It is shown here that the same conclusions hold for the current sample.
A model of the background was
obtained from its known properties and
the {\it in situ} measurements, and
the subtracted cluster fluxes remain in agreement with
those reported in our discovery papers.  They are also consistent
with results from the most conservative procedure of 
direct point-to-point
subtraction of the {\it in situ} background and proper
error propagation, which still preserves the existence of the
EUV excess and its rising radial trend.
We present
evidence which argues against the soft excess as due to peculiarities
in the line-of-sight Galactic absorption.
The data appear to favor
a thermal origin of the emission.
\end{abstract}

\section{Introduction}
Clusters of galaxies are sources which emit
EUV and soft X--ray radiation ($\sim$ 0.1-- 0.4 keV) 
substantially in excess of that expected from the hot intra--cluster medium 
(ICM) at X--ray temperatures (a few keV).
The {\it Extreme Ultraviolet Explorer} mission (Bowyer and Malina 1991) 
was pivotal to the discovery of this 
phenomenon, as it opened a unique window to the
EUV sky with the Deep Survey (DS) Lex/B filter
($\sim$ 65--190 eV, hereafter the EUV band). 
Cluster soft--excess (CSE) emission 
was reported for the Virgo (Lieu et al. 1996a), Coma (Lieu et al. 1996b), 
A1795
(Mittaz, Lieu and Lockman 1998) and A2199 clusters (Lieu, Bonamente and
Mittaz 1999).
Recent attributions of these findings 
to spatial variations of the background 
(Bowyer, Berghoefer and Korpela 1999) were based upon an incorrect 
way of analyzing the EUVE  data, and the decisive way 
of assessing the existence of CSE emission is that of an {\it in situ} 
background measurement by pointing time contiguously at small offsets
from the cluster.
This approach,  applied to a recent EUVE  re-observation 
of A2199 (Lieu et al. 1999b), clearly demonstrated
that the offset pointing provides an accurate background template for
the cluster observation, whereas an
arbitrary blank field does not.  With the aid of new EUVE observations,
performed along with acquisition of {\it in situ} background,
we were able to confirm the previous detection of CSE even in
the most conservative case when we assumed no {\it a priori}
understanding of the background distribution, so that a point-to-point
subtraction of the background had to be
adopted.  
We provide in this analysis two ways of utilizing the
background data, one is to develop a model of the background
and the other is direct subtraction of the offset field as in
the case of A2199.

\section{Virgo}
A 46 ks EUVE observation of the Virgo cluster took place in February 1999,
featuring a 34 ks pointing to the central galaxy M87 {\it immediately} 
followed by a 12 ks pointing
at $\sim$ 2 degrees offset and with the same celestial orientation of
the Lex/B filter.
This observational strategy (see Lieu et al. 1999b for details)
revealed the obvious detection of 
a halo of EUV emission around M87 of radius $\sim$  15' 
(Fig.1, top), confirming the original discovery
(Lieu et al. 1996a). 
For both source and offset fields only the central, mildly vignetted
parts of the DS Lex/B filter were considered 
\footnote{Events with Y detector coordinate in the $\sim$ 850-1200 range (Lieu et al. 1999c).
This portion of the detector Y axis covers a linear size of $\sim$ 26 arcmin;
this implies that for radii $\geq \sim$ 13 arcmin the radial profiles
are averaged only over those azimuthal angles 
that fall within the given Y axis detector limits.}.
The {\it in situ} background measurement (Fig. 1, bottom)
with the 
center of the annuli placed at the same detector position as 
that of the cluster's radial profile, indicates a
distribution of surface brightness
consistent with uniformity: the percentage deviation
between the region occupied by the cluster and that beyond
is statistically insignificant
(the difference between the count rate of the 0'--17'
and 17'--27' regions is 0.8$\pm$1.2 \%).  

To account
for any trends in the background radial profile which may be masked by
statistical fluctuations of the measured data, we utilize the
well established fact that the background
consists of a spatially uniform particle component
and a photon component which is vignetted 
(see section 4.2 and Fig. 7 of Sirk et al. 1997).  Although
the precise relative proportion varies with time, the former
usually dominates, resulting in a radial profile which is flat
or slowly varying with no change in behavior at any particular
radius, small scale ripples in the profile
are negligible (Lieu et al. 1999c).
Thus an empirical model of the background which involves a linear slope
was satisfactorily fitted to
the radial profile data at arcmin resolution.
This model is then
subtracted from
the cluster data at same resolution with propagation of errors.
Such a procedure was adopted, along with our constraints on the
hot ICM parameters (see below), to
obtain the radial trend of the fractional EUV excess, the dashed diamonds in
Fig. 2.

For the more cautious reader, however, we also provided in
Fig. 2 (dotted diamonds) 
the fractional EUV excess profile obtained from
point-to-point subtraction of the {\it in situ}
background and proper error propagation,
when the subtraction was done after scaling away a small
difference ($\sim 0.3 \%$) in the absolute background level between the
cluster and offset fields, by normalizing the 17' -- 27'
brightness of the two pointings (see Fig. 1).
The small difference between the aforementioned background
levels gives further indication that a contemporaneous, {\it in situ}
offset pointing affords the most relevant background measurement
(background levels in the DS detector have a dynamical range of
$\sim 300$ \%). Changes in the pattern
of the background distribution is a higher order effect 
than changes in the absolute level of the background, as we demonstrated
for our earlier re--observation of A2199 (Fig. 3a in Lieu et al. 1999b).
It can be seen that even
this approach, which yields the most conservative estimates of
the soft excess, indicates positive signals and a rising radial
trend.  In fact, the two subtraction procedures gave
statistically consistent results at all radii.

In order to provide an accurate model of 
the hot ICM and its emissivity in the EUV band~\footnote{The PSPC 
data used in this paper were reduced according to the prescriptions 
of Snowden et al. (1994) through the use of dedicated FTOOLS. 
Background fluxes are estimated from an outer annular region (near 45' from
PSPC boresight) where the cluster contribution is considered negligible.
Periods of high background (MV ratio 
$\geq$ 170) were also rejected and PI channels 1--19 and 202--256, which
suffer from calibration uncertainties, were 
not used.}, an archival ROSAT PSPC observation 
of the Virgo cluster (RP number
800365)
was analyzed.
Here, effects of the gradient
of Galactic line-of-sight HI column density ($N_H$) over the Virgo region,
which are not entirely negligible, need to be evaluated.
The $N_H$
was shown by a recent 21 cm measurement (Lieu et al. 1996a) to radially increase from
N$_H =$ 1.8 $\times$ 10$^{20}$ cm$^{-2}$ at the cluster center to
N$_H =$ 2.0 $\times$ 10$^{20}$ cm$^{-2}$ at a radius of 
 $\sim$ 17'.
This was confirmed by IRAS 100 $\mu$m images (Wheelock et al. 
1994; Snowden 2000), and we consequently applied 
a higher $N_H$ value to the outer regions of Virgo
when modelling the spectral data.
Given the radial HI variation (which continues its rising trend beyond
17'), and the known anti-correlation between HI and the 1/4 keV
diffuse sky background (Snowden et al. 1998), one must assess 
how much the PSPC background was underestimated for the
cluster region when it was 
determined, as in our case, from a $\sim$ 40'--50' annulus 
centered at M87.  
Of most concern is the 10'--15' annulus,
where the 1/4 keV sky background accounts for 12\% of the detected 
flux in this band.  A re--scaling of this background in accordance with
the HI gradient (Snowden et al. 1998)  over the annulus
only leads to a negligible effect on the 1/4 keV excess 
(viz. a reduction by 1\% from our reported precentage
excess, see below).
 
The emission from the hot ICM is at sufficiently low
temperature (1--3 keV) to enable
a spatially resolved temperature determination with PSPC data.
PSPC R27 band X--ray spectra  (PI channels 20--201, $\sim$ 0.2--2.0 
keV by photon energy; Snowden et al. 1994) were modelled, 
using the XSPEC software, 
with a photoelectrically absorbed single temperature
MEKAL code (Mewe, Gronenschild and Van den Oord 1985; 
Mewe, Lemen and van den Oord 1986;
Kaastra 1992), and the temperature 
and elemental abundances were fitted to the data.
By means of the hot ICM parameters
we assessed the existence of
soft--excess in the EUV measurements.  The
results, as shown in Fig. 2, are positive.  Further, they
indicate the soft excess radial trend (SERT), or more precisely
an increase in the fractional excess with radius.
It should be
mentioned that significant soft excess is also apparent in the
C-band (synonimous with R2)
 as single temperature models do not fit well the PSPC data
(Fig. 2, crosses; see also Lieu et al 1996a, Buote 2000).

\section{A1795}
A1795 was recently observed by EUVE for $\sim$ 78 ks of exposure.
The observational strategy utilizes the elongation 
of the DS Lexan/B filter (see Fig. 7 of Sirk et al. 1997).
For about half of the 
observation, cluster emission was collected on one side of the 
detector, leaving the other side 
exposed only to background counts.
For the remaining half of the time the role of the two sides
was exchanged.  Since our
calculations based on the detected
cluster EUV flux
indicate that
cluster's presence on the far side has no effect on the
asymptotic background, this
technique affords a
time contiguous measurement of the cluster signals and
the {\it in situ} 
background with equal exposures.
The background radial
profile was again found to be consistent with uniformity,
as shown in Fig. 3, where the
difference between the average count rate of the 2'--11' 
and 11'--23' regions is 1.05 $\pm$ 0.97 \%.  
The CSE effect was detected out to a radial distance of
$\sim$ 10 arcmin from the center, confirming the previous
EUVE observation (Mittaz, Lieu and Lockman 1998).
%The two possible procedures of determining and subtracting this background,
%adopted below
%to obtain statistically consistent fractional
%EUV excess profiles, were already described in section 2.

Two archival ROSAT PSPC observations (RP numbers 800055 and 800105)
were analyzed, to constrain the hot ICM parameters.
The line-of-sight HI
is smoothly distributed at $\rm N_H$ = 1.05 
$\times 10^{20}$ cm$^{-2}$ in the direction of A1795
(Mittaz, Lieu and Lockman 1998), and the R27 band spectra of annular 
regions were fitted with a MEKAL model, the abundance  being
fixed at 0.31 solar (Fabian et al. 1994).
The resulting EUV excess emission is shown in Fig. 4;
like Virgo, it exhibits
the SERT effect.  We also found  existence of CSE in the C-band, see Fig. 4 (crosses).

\section{Photoelectric absorption issues}

Throughout the analyses of this paper, we employed the absorption
cross-sections of Morrison and McCammon (1982; MM82),  whose He photoelectric
cross-section is in good agreement
with the compilation of Yan et al. (1998; Y98), the latter was
employed in a recent investigation
of the CSE
(Arabadjis and Bregman 1999).  
In the C-band the MM82 and Y98 He cross-sections
differ by $\sim$ 1\%, whereas in the
Lex/B passband ($\sim$ 65-190 eV) the Y98 cross-section is actually
{\it lower} than in MM82 by $\sim$8 \%, i.e. the new cross-section
renders our CSE fluxes even more prominent.
Given that the cross section is
not an issue, 
the proposition that the CSE detection of
some clusters (including A1795 and Virgo) is in some way
related to peculiarities in the Galactic
absorption
faces two problems.
Firstly, the $N_H$ values needed to remove the CSE
are grossly at odds with the ones measured by the dedicated 
observations of Mittaz, Lieu, \& Lockman (1998).
For example,  we found that the spectrum of the 7'--10' 
annulus of Virgo can
satisfactorily be fitted by a one temperature 
plasma model (with {\it no} soft component)
only by adopting an
$N_H$ as low as $1.34 \pm 0.07$ $\times 10^{20}$ cm$^{-2}$.
Secondly the SERT effect
(e.g., Fig. 2 and Fig. 4) implies a curious spatial correlation between
our Galactic interstellar medium and
the centers of the two clusters.
We therefore conclude that the CSE is a genuinely celestial
emission component.

\section{Interpretation of the {\it soft} 
component and conclusions}

We now discuss the physical significance of the soft component.
Currently there are two contesting models:
the non--thermal scenario, which invokes an Inverse Compton (IC) 
effect as cause for the emission (Sarazin and Lieu, 1998; Hwang 1997; Ensslin
and Biermann 1998),
and the original thermal model, which explains the CSE as thermal emission 
from a `warm' gas ($\sim 10^6$ K; Lieu et al. 1996a).

A recent study of the Coma cluster (Lieu at al. 1999a) showed 
that the CSE emission could be explained as 
an IC effect, albeit invoking a large population of energetic electrons.
When the present Virgo soft excess
was fitted with a power-law, we also 
accomodated the possibility of
`aging' due to loss of
the relativistic electrons to the ICM.~\footnote{A 
spectral code which introduces radiative (synchrotron and IC) 
and non--radiative (Coulomb) losses on a power--law model was 
integrated with the XSPEC spectral fitting package. 
As function of `aging' time (t$_{age}$) and environmental conditions
(such as density of the hot gas and magnetic field strength),
electrons at all energies are progressively removed,
modifying the resulting emissivity.}  
Even so, the fit was far from acceptable. A simple power-law model
for the soft component requires an extremely large value of the
differential photon number index (Table 1), inconsistent with
the value of $\sim$ 1.75-2 expected from a population of electrons
accelerated at astrophysical shocks (see Lieu et al. 1999a and
references therein).
When aging effects are included, a satisfactory fit can be found 
only for some of the regions, and the model was rejected for 
regions of strongest signal (0'--7' from M87; see Table 1).
The CSE can be explained, however, as thermal emission 
from a thin plasma at lower temperature ($\sim$ 0.1 keV, 
see Table 1), following the original interpretation of Lieu et al. (1996a).

Fitting of the CSE of A1795 with a non-thermal model 
produced a similar outcome.
At large radii, the EUV excess is
accompanied by a weaker C--band excess.
This necessitates a sharp cut--off, which can be understood 
as due to aging, and a highly 
evolved IC spectrum is a possible model, Table 2.  However,
the energetic requirements are severe, see Table 3, because the electron
pressure alone (in the absence of relativistic ions, which accentuates
the difficulty $\sim$ 10 - 100 times)
often exceeds that of the hot ICM.
The 6--10 arcmin region requires
relativistic electrons which,
at time of injection, had a pressure {\it $\sim$ four times higher} 
than that of the hot gas, leading to major confinement problems.
The thermal scenario again  appears more appropriate: the data of
Table 2 indicate that a $\sim$ 0.05--0.1 keV gas can satisfactorily 
explain the
CSE.

What are, then, the implications of the thermal 
interpretation on cluster masses? 
In absence of any knowledge on the clumping of the warm phase, a 100 \% 
filling factor implies a warm gas mass comparable with that of the hot gas
(Table 4).  This is in 
good overall agreement with recent large--scale hydrodynamical simulations
(Cen and Ostriker 1999) which indicate that at the present epoch the baryons 
are to be found, in similar amounts, at hot ($\sim 10^8$ K) and warm 
($\sim 10^5-10^7$ K) temperatures,
and only marginally at lower temperatures (see also Bonamente,
Lieu and Mittaz 2000 for the detection of cold clouds
with commensurately smaller mass budgets).

\newpage
{\bf References} \\

\noindent Arabadjis,J.S. and Bregman, J.N. 1999, {\it ApJ}, {\bf 514}, 607. \\
\noindent Balucinska-Church, M. and McCammon, D. 1992, {\it ApJ}, {\bf 400}, 699. \\
\noindent Bonamente, M., Lieu, R. and Mittaz, J.P.D. 2000, {\it ApJ in press}. \\ 
\noindent Bowyer, S. and Malina, R.F. 1991, in {\it EUV Astronomy}, \\
\indent eds. R.F. Malina and S. Bowyer (New York:Pergamon), 397. \\
\noindent Bowyer, S., Bergh\"{o}fer, T. and Korpela, E. 1999, {\it ApJ}, {\bf 526}, 592. \\
\noindent Briel, U.G. and Henry, J.P. 1996, {\it ApJ}, {\bf 472}, 131.\\
\noindent Buote, D., 2000, {\it ApJ in press}. \\
\noindent
Cen, R. and Ostriker, J.P. 1999, {\it ApJ}, {\bf 514}, 1. \\
\noindent Ensslin, T.A. and Bierman, P.L. 1998, {\it A\&A}, {\bf 330}, 90.\\
\noindent Fabian, A.C., Arnaud, K.A., Bautz, M.W. and Tawara, Y. 1994, \\
\indent {\it ApJ}, {\bf 436}, L63. \\ 
\noindent Hwang, C.-Y. 1997, {\it Science}, {\bf 278}, 1917. \\
\noindent
~Kaastra, J.S. 1992 in \it An X-Ray Spectral Code for Optically Thin Plasmas \rm
 \\\indent
(Internal SRON-Leiden Report, updated version 2.0). \\
\noindent
Lieu, R., Mittaz, J.P.D., Bowyer, S., Lockman, F.J., Hwang, C.-Y. and \\
\indent Schmitt, J.H.H.M. 1996a, ApJ, {\bf 458}, L5.\\
\noindent
 Lieu, R., Mittaz, J.P.D., Bowyer, S., Breen, J.O.,
Lockman, F.J., \\
\indent Murphy, E.M. \& Hwang, C. -Y. 1996b, {\it Science}, {\bf 274}, 1335. \\
\noindent Lieu, R., Bonamente, M. and Mittaz, J.P.D. 1999, 
{\it ApJ}, {\bf 517}, L91.\\
\noindent Lieu, R., Ip, W.-I., Axford, W.I. and Bonamente, M. 1999a, {\it ApJ}, 
{\bf 510}, L25.\\
\noindent
Lieu, R., Bonamente, M., Mittaz, J.P.D., Durret, F., Dos Santos, S. and \\
\indent Kaastra, J. 1999b, ApJ, {\bf 527}, L77.\\
\noindent
Lieu, R., Bonamente, M., Mittaz, J.P.D., Durret, F., and Dos Santos, S. 1999c, \\
\indent
{\it Proceedings of the Workshop on 'Diffuse Thermal and Relativistic Plasma in \\
\indent
Galaxy Clusters'}, Ringberg, Germany MPE report 271, p.197 \\
\indent  (http://www.xray.mpe.mpg.de/theorie/cluster/ringb99\_proc.html). \\ 
\noindent
~Mewe, R., Gronenschild, E.H.B.M., and van den Oord, G.H.J., 1985 
 {\it A\&A Supp.}, {\bf 62}, 197.  \\
\noindent Mewe, R., Lemen, J.R., and van den Oord, G.H.J. 1986,
\it A\&A. Supp.\rm, \\
\indent {\bf 65}, 511--536. \\
\noindent
~Mittaz, J.P.D., Lieu, R., Lockman, F.J. 1998, {\it ApJ}, {\bf
498},
L17. \\
\noindent
 Morrison, R. and McCammon, D. 1983, {\it ApJ}, {\bf 270}, 119.\\
\noindent Nulsen, P.E.J. and B\"{o}hringer, H. 1995, {\it MNRAS}, {\bf 274}, 1093.\\
\noindent Sarazin, C.L. and Lieu, R. 1998, {\it ApJ}, {\bf 494}, L177. \\
\noindent Sirk, M.M., Vallerga, J.V., Finley, D.S., Jelinsky, P.,
and Malina, R.F. 1997, \\
\indent {\it ApJS}, {\bf 110}, 347. \\
\noindent Snowden, S.L., McCammon, D., Burrows, D.N. and Mendenhall, J.A. 1994, 
 {\it ApJ}, {\bf 424}, 714. \\
\noindent Snowden, S.L., Egger, R., Finkbeiner D.P., Freyberg, M.J. and  \\
\indent Plucinsky, P.P. 1998, {\it ApJ}, {\bf 493}, 715.\\
\noindent Snowden 2000, in preparation. \\
\noindent Wheelock et al. 1994, {\it IRAS Sky Survey Explanatory Supplement}, \\
\indent (JPL Publication 94-11), Pasadena, CA. \\
\noindent Yan, M., Sadeghpour, H.R. and Dalgarno, A. 1998, {\it ApJ}, {\bf 496}, 1044.\\

\newpage
{\bf Figure captions}

Figure 1: Radial profile of the cluster 
pointing (top) and offset pointing (bottom) of the Virgo
observation. 
For both plots the radius is measured from the detector position of M87.
The innermost 1' bin 
of the cluster, dominated by the emission from M87, is omitted.
The dotted   
line represents average brightness of the 17'--27' region.
On same ordinate of the cluster pointing (top) is overplotted the radial
profile of the offset pointing with open circles, 
after a $\sim$ 0.3 \% rescaling (the difference
between the 17'--27' region brightness of the two profiles) is applied.
A radial profile of offset pointing (bottom) in 1 arcmin resolution was
successfully fitted to an empirical linear slope, and its best-fit parameters
used for the evaluation of the fractional EUV excess (dashed diamonds
in Fig. 2; see also text).

Figure 2: Diamonds are radial profile of the EUV fractional excess 
$\eta$ of Virgo, defined as
$\eta = (p-q)/q$, where $p$ is the observed EUV band flux after 
background subtraction,
and $q$ is the expected flux from the hot ICM.
Vertical semi-diameters are here and after 1-$\sigma$ errors.
Dashed diamonds are obtained by subtraction of model background,
dotted diamonds by point to point background subtraction. 
Crosses are 
radial profile of the C-band fractional excess $\eta$ of Virgo.

Figure 3: Radial profile of the co-added cluster pointings (top) 
and offset pointings (bottom) of the A1795  
observation.  The innermost 2' bins are omitted, 
as there was substantial contamination from a
bright unidentified transient source
near the cluster center.
On either plot the
dotted
line represents average brightness of the 11'--23' region.
On same ordinate of the cluster pointing (top) is overplotted the radial
profile of the offset pointing with open circles,
after a $\sim$ 0.7 \% rescaling (the difference
between the 11'--23' region brightness of the two profiles) is applied.
A radial profile of offset pointing (bottom) in 1' resolution was
successfully fitted to an empirical linear slope, and its best-fit parameters
used for the evaluation of the fractional EUV excess (dashed diamonds
in Fig. 4).

Figure 4: Diamonds are radial profile of the EUV fractional 
excess $\eta$ for A1795.  Owing to contamination from a transient source 
(see Figure 3),
data for the 0--2 arcmin bins are from 
a previous observation
of the cluster (Mittaz, Lieu and Lockman 1998).  Note also that
this previous paper showed a similar plot with a stronger
rising trend, because it used absorption code of Balucinska-Church
and McCammon (1992), whereas the present work uses the original
Morrison \& McCammon (1982) code, which is now believed to
involve a more reliable He cross-section (see text).
Dashed diamonds are obtained by subtraction of model background,
dotted diamonds by point to point background subtraction (see section 2).
Crosses are
radial profile of C-band fractional excess of A1795.

\newpage

\begin{table}[h!]
\begin{center}
\tighten
\caption{Modelling of Virgo spectra with a two--temperature MEKAL code (left), 
MEKAL + power-law
model (center) and MEKAL + `aged' power-law  model (right). Errors
are 90 \% confidence ($\chi^2$ + 2.701 criterion).
 For the latter, the IC 
magnetic field was fixed at 1$\mu$G, 
the density of the ICM calculated from the best-fit $\beta$-model
of Nulsen and B\"{o}hringer (1995) and
the electron differential number index $\gamma$ fixed at the
value of 2.5 (wich corresponds to a similar photon index $\alpha$ of 1.75).
The parameters of the hot phase were fixed at those best values obtained
by fitting PI channels 42 -- 201 with a single temperature model of
floating abundance.
$N_{warm}$ is here and after in
units of $10^{-14}$/4$\pi D^2 n^2 V$, where $D$ (cm) is the distance to the source, 
$n$ (cm$^{-3}$)
is the gas density and $V$ (cm$^3$) is the volume of the emitting region.
For MEKAL+MEKAL and MEKAL+PO model fits, reduced $\chi^2$ is always between
1.1 and 1.3 for 180--181 d.o.f..
\label{tab:virgo}}
\vspace{22pt}
\hspace{-2cm}
\small
\begin{tabular}{cccccc}
\hline
 & \multicolumn{2}{c}{MEKAL+MEKAL} & \multicolumn{1}{c}{MEKAL+PO} 
& \multicolumn{2}{c}{MEKAL+`aged' PO}\\
Region & $T_{warm}$ & $N_{warm}$ 
& $\alpha$ & $t_{age}$ & red. $\chi^2$ \\
 (arcmin) & (keV) & $\times 10^2$ & & (Gyrs) & (d.o.f.)  \\
\hline
0--3 & 0.066 $\pm^{0.038}_{0.014}$ & 0.36 $\pm^{0.24}_{0.16}$ 
& 3.3 $\pm^{2.4}_{0.6}$ & $\sim$ 0 & 1.6(181) \\
3--5 & 0.079 $\pm^{0.039}_{0.016}$ & 0.28 $\pm^{0.07}_{0.08}$ 
& 4 $\pm^{1.4}_{0.6}$ & $\sim$ 0 & 1.66(181) \\
5--7 & 0.079 $\pm^{0.022}_{0.015}$ & 0.4 $\pm^{0.2}_{0.12}$ 
& 4.4 $\pm^{1.2}_{0.7}$ & $\sim$ 1.1 & 2(180) \\
7--10& 0.09 $\pm^{0.01}_{0.012} $ & 0.59 $\pm^{0.25}_{0.13}$ 
& 4.2$\pm^{0.8}_{0.5}$ & 1.8$\pm^{0.015}_{0.02}$ & 1.11(180) \\
10--15&0.083 $\pm^{0.009}_{0.011}$ & 1.1 $\pm^{0.38}_{0.1}$ 
& 4.5$\pm^{0.65}_{0.3}$ & 2.1$\pm^{0.15}_{0.05}$ & 1.2(180) \\
\hline
\end{tabular}
\end{center}
\end{table}

\begin{table}[h!]
\begin{center}
\tighten
\caption{
Modelling of A1795 spectra with a two-temperature MEKAL code (left) and
with a MEKAL + `aged' power-law code (right);
errors are 90\%
confidence ($\chi^2$ + 2.701 criterion). Parameters of the hot phase were
fixed at those best values obtained by fitting PI channels 42 -- 201 with
a single temperature model of abundance 0.31 solar (Fabian et al 1994).
For the non-thermal model, the value of the 
IC magnetic field was fixed at 1$\mu$G; the density of the ICM, a parameter which
affects the aging of electrons, was calculated from the best-fit
$\beta$-model of Briel and Henry (1996). Electron differential index $\gamma$ was fixed 
at the value of 2.5 (see Table 1).
\label{tab:a1795_non}}
\vspace{22pt}
\small
\begin{tabular}{ccccccc}
\hline
 & \multicolumn{3}{c}{MEKAL+MEKAL} & \multicolumn{2}{c}{MEKAL+`aged' PO}  \\
% & \multicolumn{3}{c}{} & \multicolumn{2}{c}{} \\
Region  & $T_{ warm}$ & $N_{warm}$ &red. $\chi ^2$ 
  & $t_{age}$  & Red. $\chi^2$ \\
 (arcmin) & (keV)&$\times 10^2$ & (d.o.f.) &  (Gyrs)    & (d.o.f.)  \\ 
\hline
1--2 & 0.125$\pm^{0.034}_{0.037}$ & 0.045$\pm^{0.008}_{0.009}$ & 1.0(363)
& 1.7 $\pm^{0.17}_{0.11}$ &   1.0(363)\\
2--3 & 0.11$\pm^{0.04}_{0.02}$& 0.0285$\pm^{0.006}_{0.007}$ & 1.16(363)
& 2.25 $\pm^{0.15}_{0.55}$  & 1.15(363)\\
3--6 & 0.049$\pm^{0.017}_{0.015}$& 0.12$\pm^{0.37}_{0.06}$ & 0.97(363) 
& 2.8 $\pm^{0.1}_{0.13}$ & 0.98(363) \\
6--10& 0.025$\pm^{0.013}_{0.012}$ & 110$\pm^{700}_{70}$ & 0.88(363)
& 3.7 $\pm^{0.16}_{0.14}$    &  0.88(363) \\
\hline
\end{tabular}
\end{center}
\end{table}

\begin{table}[h!]
\begin{center}
\tighten
\caption{Luminosity and pressure estimates for A1795;
$L^{CSE}_{42}$ is the intrinsic luminosity of non--thermal best--fit
model between photon energies 65--250 eV ($\gamma_{min}$=279 and
$\gamma_{max}$=548) in units of 10$^{42}$ (erg s$^{-1}$) and
$P^{CSE}$ is the pressure of relativistic electrons
in the [$\gamma_{min}$,$\gamma_{max}$] Lorentz factor interval.
$P^{e}$ and $P^e(t=0)$ are
the total electron pressure at the present epoch
and at the acceleration epoch, respectively, assuming cosmological
parameters $H_0=75$ km s$^{-1}$ Mpc$^{-1}$ and $q_0=0$. If $H_0=50$ km s$^{-1}$ Mpc$^{-1}$
is adopted, all present pressure estimates for the 
non-thermal component will increase $\sim$ two-fold.
Pressure for the ICM ($P^{gas}$) was calculated according
to the parameters of Briel and Henry (1996).
\label{tab:a1795_pr}}
\vspace{22pt}
\small
\begin{tabular}{cccccc}
\hline
Region & $L^{CSE}_{42}$ & $P^{CSE}$ & $P^{gas}/P^{CSE}$ &
$P^{gas}/P^e$ & $P^{gas}/P^e(t=0)$ \\
(arcmin) & & (erg cm$^{-3}$) & & &  \\
\hline
1--2 & 0.9 & 1.6 $10^{-13}$ & 88.3 & 35.3 & 1.9  \\
2--3 & 1.3 & 1.0 $10^{-13}$ & 128& 85.8 & 4.8  \\
3--6 & 4.5 & 1.0 $10^{-13}$ & 120 & 52.7 & 2.3   \\
6--10& 22.0& 6.8 $10^{-14}$ &  17.6& 2.5  & 0.22  \\
\hline
\end{tabular}
\end{center}
\end{table}

\begin{table}[h]
\begin{center}
\tighten
\caption{Warm gas mass estimates and comparison with the hot ICM for
the Virgo and A1795 cluster (see Tables 1 and 2). Warm gas densities ($n_{ warm}$)
are estimated for 100\% filling factors.}
\vspace{22pt}
\small
\begin{tabular}{cccccc}
\hline
\multicolumn{3}{c}{Virgo} & \multicolumn{3}{c}{A1795} \\
%\multicolumn{3}{c}{--------} & \multicolumn{3}{c}{--------} \\
Region & $n_{ warm}$ & $
n_{hot}$ & Region & $n_{warm}$ & $n_{hot}$\\
(arcmin) & ($10^{-3}$ cm$^{-3}$) &($10^{-3}$ cm$^{-3}$) 
& (arcmin) & ($10^{-3}$ cm$^{-3}$) & ($10^{-3}$ cm$^{-3}$) \\
\hline
0--3 & 5.2   $\pm^{3.4}_{2.4}$     & 50 & 0--1 & --- & 3.7 \\
3--5 & 2.4 $\pm^{0.6}_{0.65}$   &10  & 1--2 & 0.92$\pm^{0.16}_{0.18}$ & 3.3 \\
5--7 & 1.9 $\pm^{1}_{0.6}$   &7.5 & 2--3 & 0.45$\pm^{0.1}_{0.11}$  & 2.7 \\
7--10& 1.4 $\pm^{0.6}_{0.3}$   &5   & 3--6 & 0.3$\pm^{0.9}_{0.15}$  & 1.6 \\
10--15&1.0$\pm^{0.34}_{0.1}$ &3   & 6--10& 4.5$\pm^{26}_{2.6}$     & 0.93\\
\hline
\end{tabular}
\end{center}
\end{table}

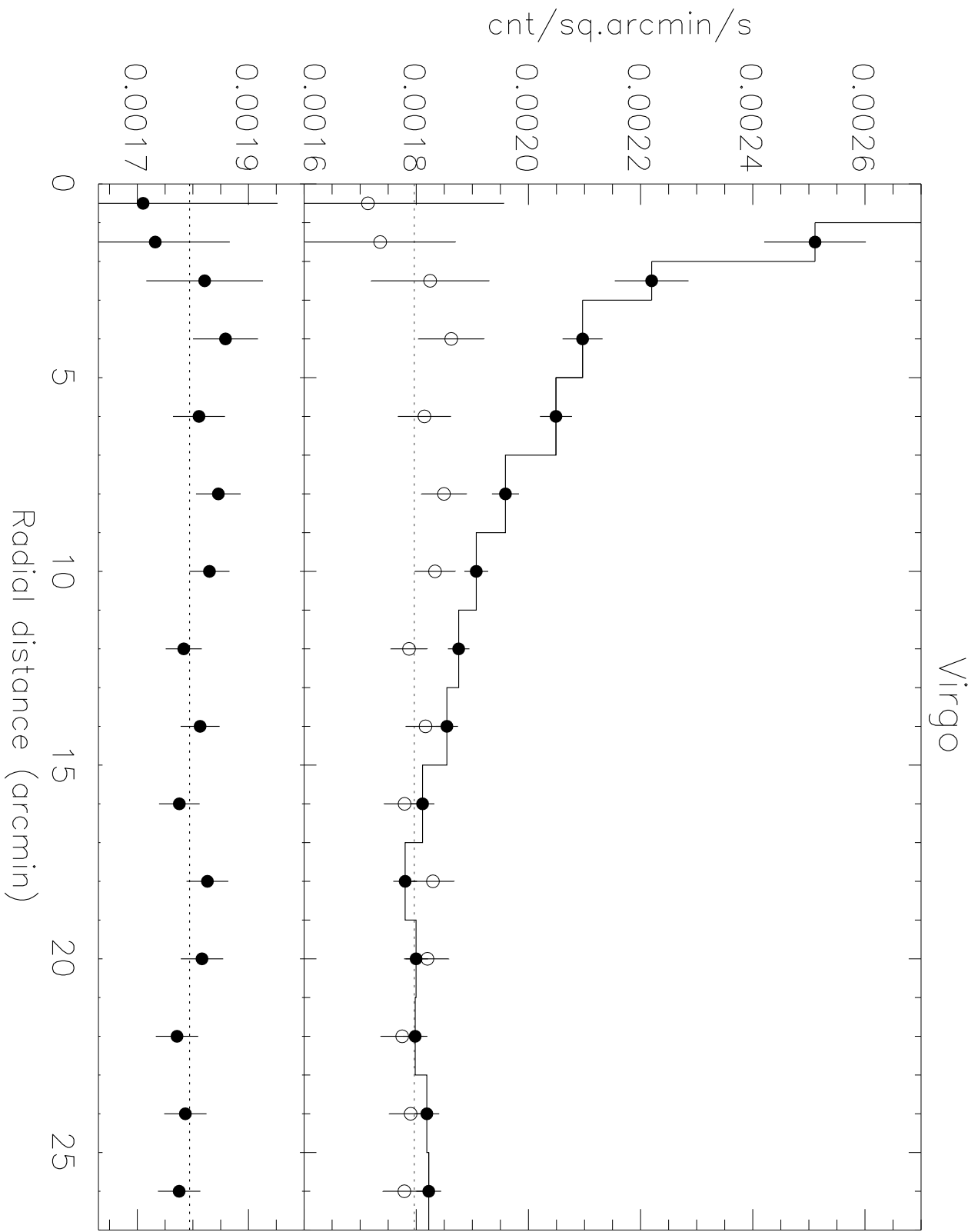
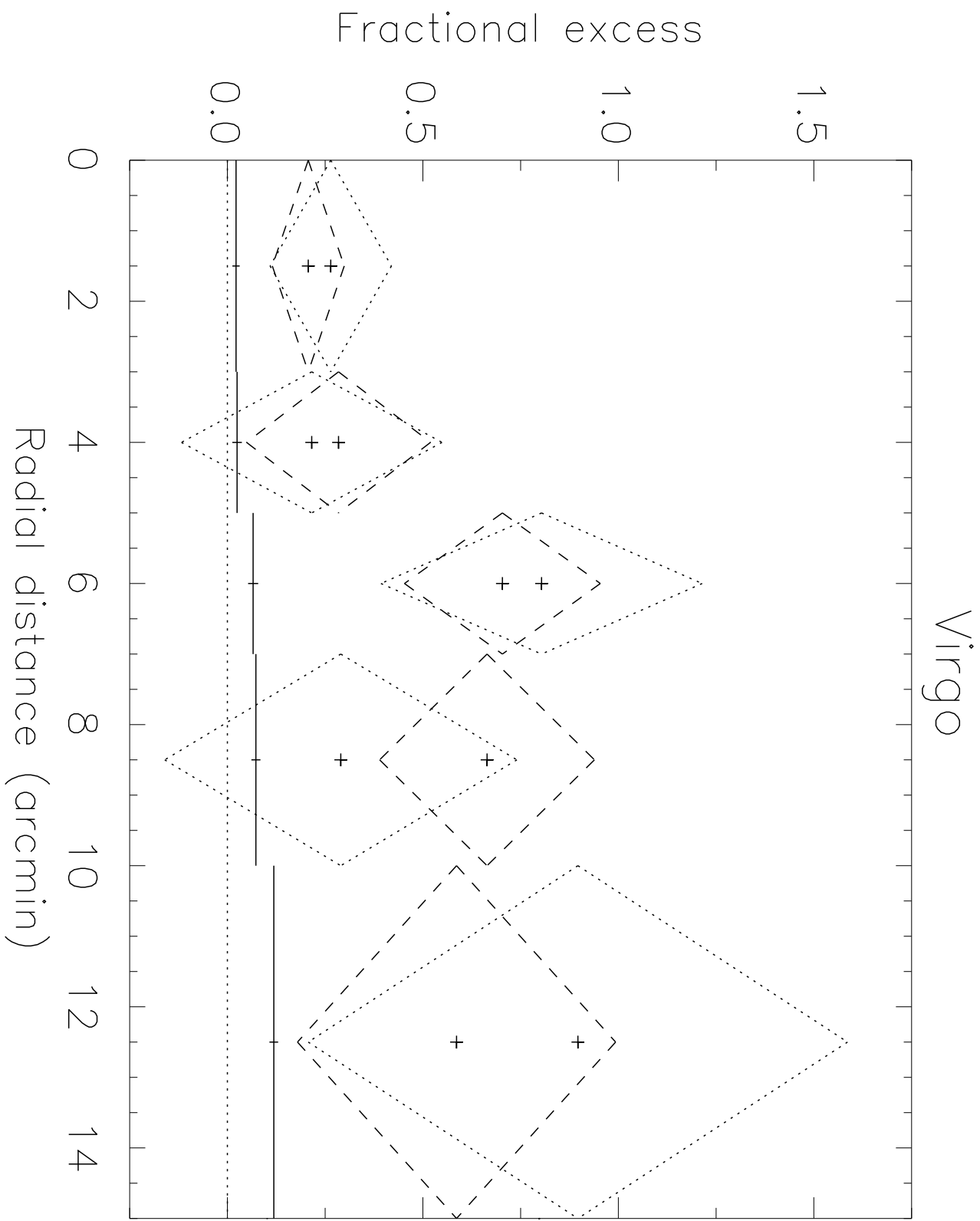
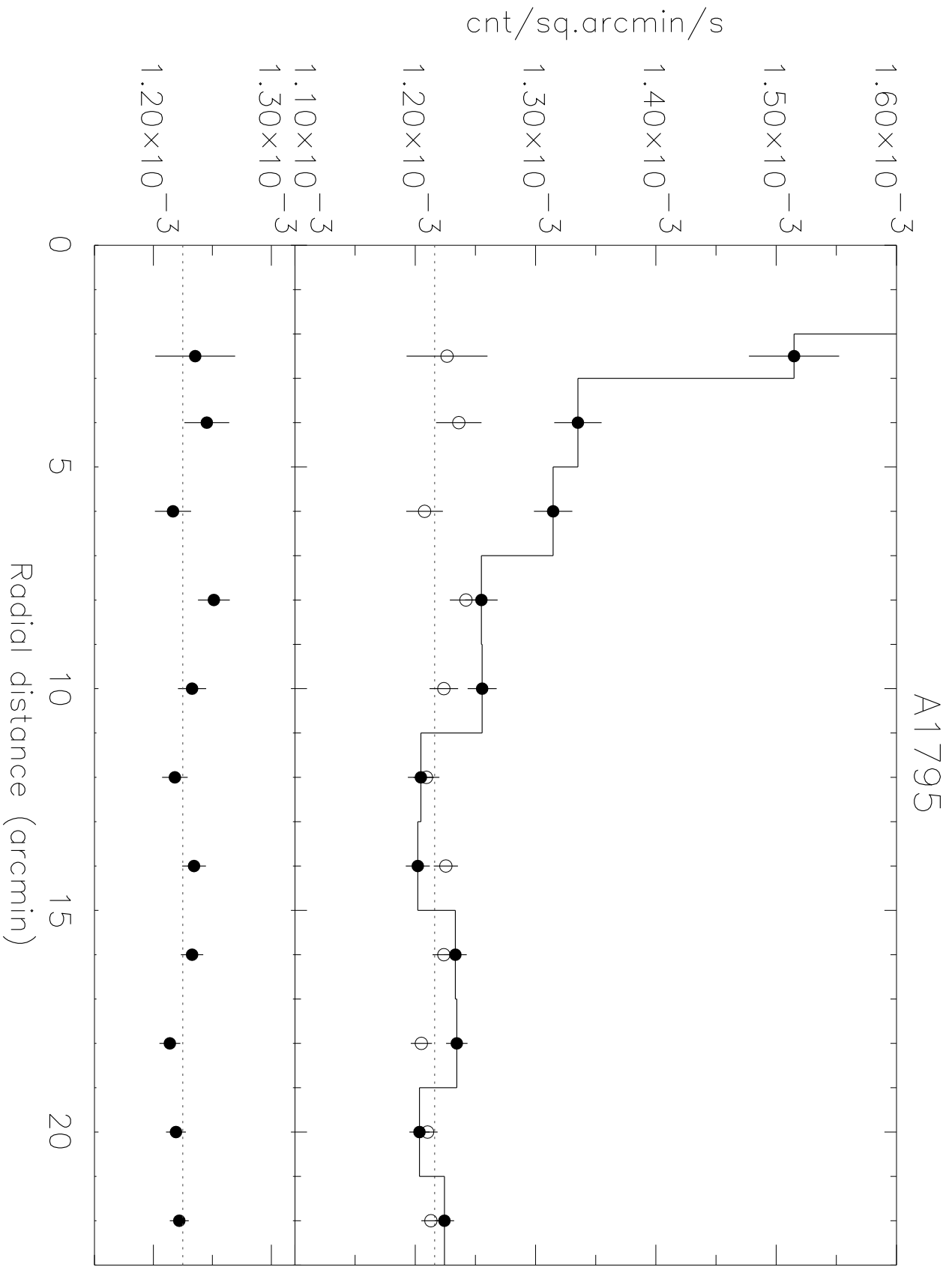
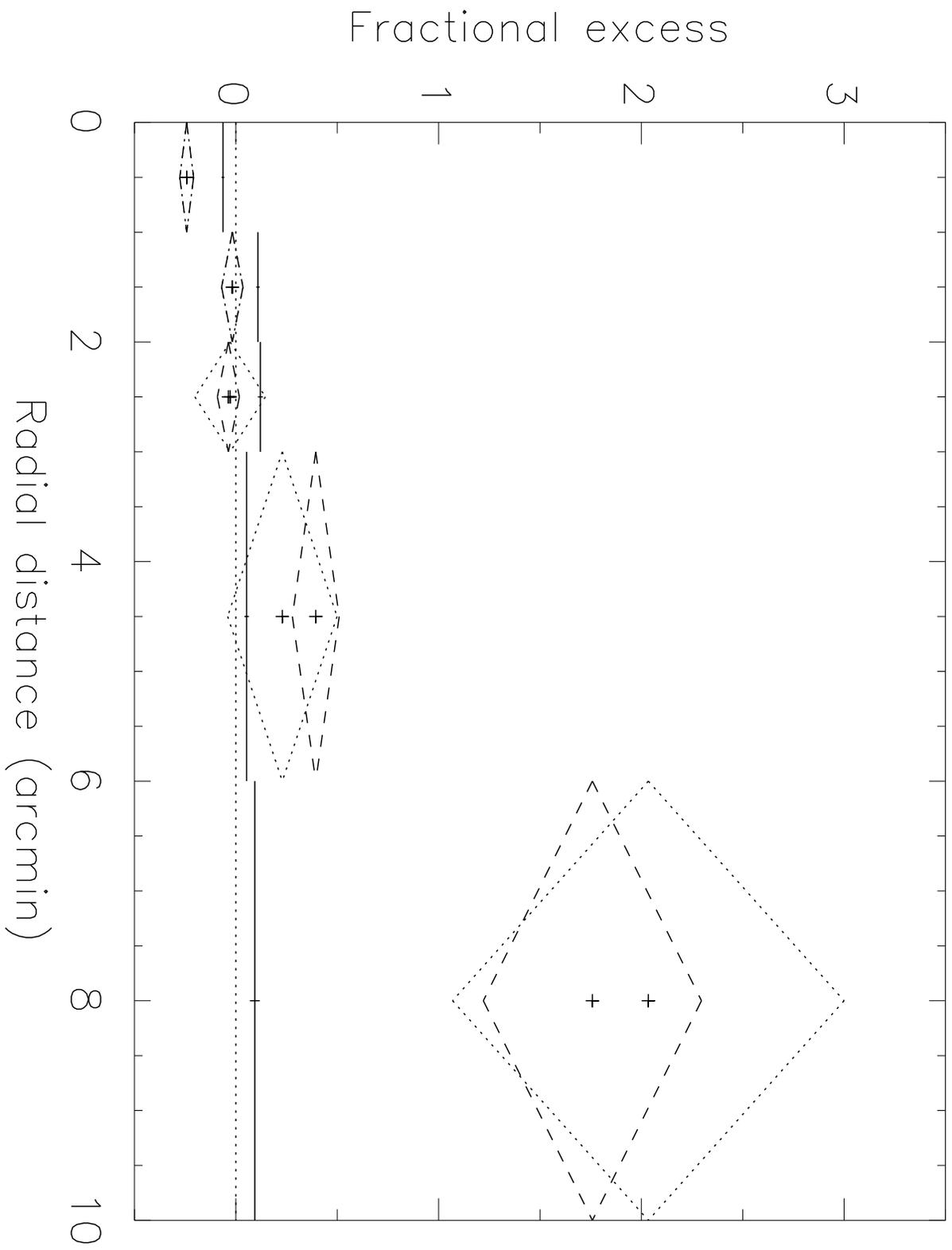
\end{document}